\documentclass[conference]{IEEEtran}
\IEEEoverridecommandlockouts
\usepackage{cite}
\usepackage{amsmath,amssymb,amsfonts}
\usepackage{algorithmic}
\usepackage{graphicx}
\usepackage{textcomp}
\usepackage{xcolor}
\def\BibTeX{{\rm B\kern-.05em{\sc i\kern-.025em b}\kern-.08em
    T\kern-.1667em\lower.7ex\hbox{E}\kern-.125emX}}

\begin{document}

\title{Demonstration of a Networked Music Performance Experience with MEVO
}

\author{
\IEEEauthorblockN{Leonardo Severi, Matteo Sacchetto, Andrea Bianco, Cristina Rottondi}
\IEEEauthorblockA{\textit{Dept. of Electronics and Telecommunications} \\
\textit{Politecnico di Torino}\\
Turin, Italy \\
\{name.surname\}@polito.it}
\and
\IEEEauthorblockN{Aleksandra Knapińska, Piotr Lechowicz}
\IEEEauthorblockA{\textit{Dept. of Systems and Computer Networks} \\
\textit{Wrocław University of Science and Technology}\\
Wrocław, Poland \\
\{name.surname\}@pwr.edu.pl}

}

\maketitle

\begin{abstract}
In this paper we present a Networked Music Performance system currently under development at Politecnico di Torino. We demonstrate its use in a distributed concert held in June 2023, which featured three musicians in Turin (Italy) and three musicians in Wrocław (Poland). Although in its early stages, the system proved to be already stable enough to appear transparent to the remote audience. 
\end{abstract}


\section{Introduction} \label{sec:introduction}
A Networked Music Performance (NMP) refers to a live, real-time interaction between musicians mediated by a telecommunication network. 
An NMP framework is a system providing the technological facilities to support remote musical performances. 
The objective of a NMP system is to be as transparent as possible to its users, enabling them to play together as if they were in the same room. 

In this paper we present an example of experience with MEVO (Music rEVOlution), the NMP system currently under development at Politecnico di Torino, by describing its use in a remote concert which took place in June 2023 and involved performers in two locations: Politecnico di Torino and Wrocław University of Science and Technology. 

The remainder of the paper is structured as follows: section \ref{sec:background} provides a brief background on NMP, section \ref{sec:experiment} presents the experience of the distributed concert and section \ref{sec:results} shortly discusses the results based on telemetry data collected during the experiment.

\section{Background} \label{sec:background}

In NMP, the key aspect to keep in consideration is the end-to-end audio latency, also called mouth-to-ear (M2E) latency. To guarantee a realistic musical interaction, ideally the M2E latency should be kept below 30 ms \cite{rottondi-overview-nmp}, though, depending on the musical piece and on the musicians' ability to apply delay-coping strategies, higher values may still be tolerated. Usually, latencies above 60ms lead to significant tempo deceleration, which degrades the musical performance \cite{bosi2021experiencing}.

The second aspect to take into account is that latency varies along time due to several reasons, e.g., evolving network traffic conditions, misalignment of audio card clocks (audio drift), etc. Thus, a NMP system should deal with these issues in a way as transparent as possible to musicians, aiming to keep the perceived latency as steady as possible. Our NMP system addresses jitter issues by relying on a dynamic estimation of the playback buffer, i.e. the buffer that stores incoming audio frames prior to their reproduction, that dynamically adapts to the network conditions.

The last issue to be addressed by a NMP system is packet loss concealment (PLC). Due to the real-time requirements of NMP, the common solution is to rely on the UDP transport protocol for audio streaming. The side effect of using UDP is that it provides no guarantees on the reliability of packet delivery, therefore packet losses may occur. In our current implementation of the system, no packet loss concealment technique is implemented, so in case of packet loss missing samples are replaced with silence, but PLC techniques such as those presented in \cite{sacchetto-ar} are in the process of being integrated.

In literature many NMP systems are available (e.g. JackTrip \cite{jacktrip}, Sonobus \cite{sonobus}, SoundJack \cite{soundjack}).  The final implementation of MEVO will differ from those already available at either commercial or experimental stage, as it will specifically focus on supporting remote music education and integrate dedicated hardware and software to ensure accessibility also by visually, auditory and mobility-impaired users.

\section{Experiment} \label{sec:experiment}
MEVO features a Raspberry Pi 4B connected to a USB audio card. 
The software part consists in the headless version of the Raspberry Pi Operating System (Raspberry Pi OS Lite), with Linux 6.1 PREEMPT\_RT, a control system written in JavaScript and running on Node.js 18.16 and the main program written in C++17. Furthermore, the operating system runs at the stock clock of $1.5GHz$
, using 3 over 4 of its cores for general purpose activities, whereas the fourth one is reserved for the management of input/output to/from the audio card. The main program is responsible for capturing the Pulse Code Modulation (PCM) audio by means of ALSA's APIs, possibly converting the sample format (without adding any extra encoding), encapsulating it in UDP packets and transmitting it to all the other players in a peer-to-peer fashion. Concurrently, it receives the packets from the other players' devices, buffers the audio samples they contain and plays them back through the sound card.

We demonstrated MEVO in a distributed concert between the cities of Wrocław (Poland) and Turin (Italy). The setup was the same in both locations. The concert featured three musicians located in Turin (a cellist, a flutist, and a pianist playing a midi keyboard), and three in Wrocław (a singer, a clarinetist and a percussionist). The live performance lasted approximately one hour, but the system was powered on and working for $\approx 2h\, 45'$. The musical pieces were composed ad hoc for the event, and premiered during the concert. The musicians opted for playing five pieces with the help of a metronome click generated at the Turin side. During this part of the concert, the musicians performing in Turin were not listening to their counterpart in Wrocław, whereas the musicians in Wrocław were receiving the metronome's click together with the live audio of the musicians in Turin. The metronome click was only audible by the musicians and not by the in presence audience at the Wrocław side.
Instead, the last piece featured the cellist and the percussionist playing without any additional synchronization mechanisms.
\begin{figure}
    \centering
    \includegraphics[width=\linewidth]{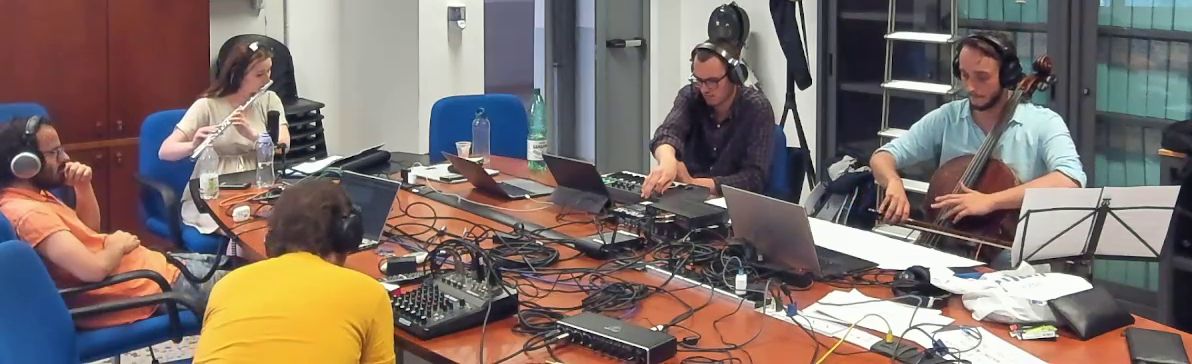}
    \vspace{-2em}
    \caption{Photography of musicians performing in Turin}
    \vspace{-1em}
    \label{fig:turin_photo}
\end{figure}

\section{Results}\label{sec:results}
\begin{figure}[tb]
    \centering
    \includegraphics[width=\linewidth]{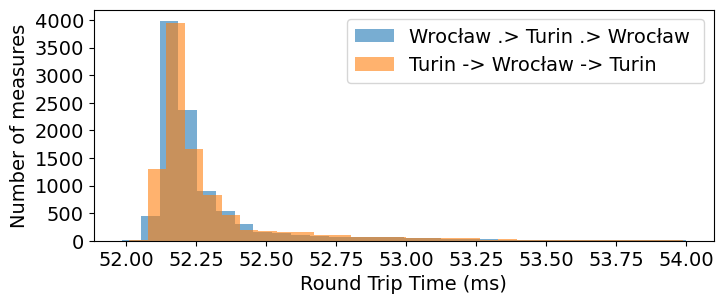}
    \vspace{-2.5em}
    \caption{Distribution of Round trip times. Figure reports value between 0 and $54ms$ for ease of read.}
     \vspace{-0.5em}
    \label{fig:rtt_hist}
\end{figure}

At both sides, we configured the NMP software to collect telemetry data, including the automatic sizing of the playback buffer, the round trip time (RTT) and the amount of lost audio frames, either due to lost UDP packets or to late ones. 
The data capture process took place regularly once every second by a software thread designed to minimize the impact on the performance of the rest of the software. The minimum measured RTT value was $51.985\,ms$ at one side and $52.011\,ms$ at the other one. Overall, the network exhibited a stable behavior, $99.986\%$ of the measured RTTs was $<59ms$ at both sides. Fig. \ref{fig:rtt_hist} shows the RTT distribution. 

\begin{figure}
    \centering
    \includegraphics[width=\linewidth]{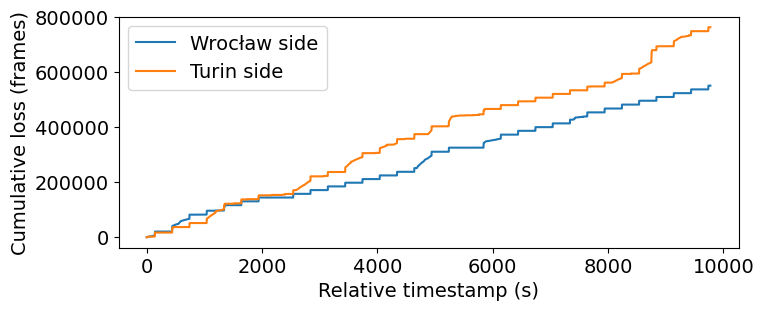}
    \vspace{-2.5em}
    \caption{Cumulative losses at both sides in terms of absolute count of lost audio frames.}
    \vspace{-0.5em}
    \label{fig:cumulative_loss}
\end{figure}

Fig. \ref{fig:cumulative_loss} illustrates the amount of lost frames during the concert. The sampling rate was $44100$ Hz, thus that the total count of lost frames at Turin side corresponds to approximately $18s$ of lost audio over $\approx 2h\, 45'$. More precisely, the ratios of lost frames are $0.131\%$ at Wrocław side and $0.177\%$ at Turin side.
Furthermore, based on measurements conducted prior to the concert, we estimated that the audio driver and soundcard latency contribution is $\approx 5ms$, and that the receiver buffer added a delay of $18.34ms$ (Confidence interval: $[1.29,29.75]ms$) at Wrocław side, and $10.88ms$ (Confidence interval: $[1.13, 30.25]ms$) at Turin side. Thus, the overall M2E latency experienced by the musicians was within $[32,61]ms$.

Even though the development of MEVO is in its early stage, the overall feedback received from the musicians and the audience was that the system is already stable and transparent enough to introduce negligible impact on the perceived quality of experience. Nevertheless, the experiment highlighted that integration of PLC techniques is needed to address packet losses, that in some cases were clearly audible, and that further improvement of the audio buffer dynamic estimation methodology is necessary.

\section*{Acknowledgments} \label{sec:ack}
This work has been supported by the Italian Ministry for University and Research under the PRIN program (grant n. 2022CZWWKP).

Leonardo Severi's PhD Program is funded by the European Union in the framework of the Resiliency and Recovery Plan (RRP), within the NextGenerationEU initiative. 

The authors thank the performers Karol Knapiński, Michał Kram, Kinga Kubiak, Karolina Kułaga, Sara Warzecha, and Olgierd Żemojtel for their participation. 

\bibliographystyle{IEEEtran} 
\bibliography{refs.bib}

\end{document}